\documentclass[twocolumn,showpacs,preprintnumbers,amsmath,amssymb,latexsym,prl,nofootinbib]{revtex4-1}
\usepackage{latexsym,amssymb,amsfonts,amsmath}
\usepackage{color}
\usepackage[dvips]{graphicx}

\newcommand{\msub}[1]{\ensuremath _{\mbox{\scriptsize #1}}}

\newcommand{\nii}{n\msub{i}}
\newcommand{\naa}{n\msub{a}}

\newcommand{\figsize}{0.48}

\usepackage{graphicx}% Include figure files
\usepackage{dcolumn}% Align table columns on decimal point
\usepackage{bm}% bold math
\usepackage{ulem}

\begin{document}

%\preprint{APS/123-QED}

\title{Fate of Chiral Symmetries in the Quark-Gluon Plasma from an
  Instanton-Based Random Matrix Model of QCD} \thanks{I thank Matteo Giordano,
  S\'andor Katz, Attila P\'asztor and D\'aniel N\'ogr\'adi for discussions,
  and Matteo Giordano for a careful reading of the manuscript. This work was
  supported by Hungarian National Research, Development and Innovation Office
  NKFIH Grant No. KKP126769 and NKFIH excellence grant TKP2021-NKTA-64.}

\author{Tam\'as G.\ Kov\'acs} 

\affiliation{%
  Department of Physics and Astronomy, ELTE E\"otv\"os Lor\'and University \\
  H-1117 Budapest, P\'azm\'any P\'eter s\'et\'any 1/a \\
  and \\
  Institute for Nuclear Research (ATOMKI), \\
  H-4026 Debrecen, Bem t\'er 18/c, Hungary %\textbackslash\textbackslash
}%

\date{\today}% It is always \today, today,

\begin{abstract}  
  We propose a new way of understanding how chiral symmetry is realized in the
  high temperature phase of QCD. Based on the finding that a simple free
  instanton gas precisely describes the details of the lowest part of the
  spectrum of the lattice overlap Dirac operator, we propose an
  instanton-based random matrix model of QCD with dynamical
  quarks. Simulations of this model reveal that even for small quark mass the
  Dirac spectral density has a singularity at the origin, caused by a dilute
  gas of free instantons. Even though the interaction, mediated by light
  dynamical quarks creates small instanton-antiinstanton molecules, those do
  not influence the singular part of the spectrum, and this singular part is
  shown to dominate Banks-Casher type sums in the chiral limit. By
  generalizing the Banks-Casher formula for the singular spectrum, we show
  that in the chiral limit the chiral condensate vanishes if there are at
  least two massless flavors. Our model also indicates a possible way of
  resolving a long-standing debate, as it suggests that for two massless quark
  flavors the $U(1)\msub{A}$ symmetry is likely to remain broken up to
  arbitrarily high finite temperatures.
\end{abstract}

%\pacs{12.38.Gc,72.15Rn,12.38.Mh,11.15.Ha}% PACS, the Physics and Astronomy
\maketitle

{\it Introduction.---} Quantum chromodynamics (QCD), the theory of strong
interactions has been around for 50 years, but we still have not explored all
of its consequences. One of the outstanding problems is the lack of a full
understanding of the finite temperature transition from hadrons to the
quark-gluon plasma, and in particular the restoration of the spontaneously
broken chiral symmetry, and the loss of quark confinement.

The lightest quarks in nature, the $u$ and $d$, making up most of the visible
matter around us, have much smaller mass than the characteristic scale of
QCD. As a result, the $SU(2)\msub{A} \times U(1)\msub{A}$ chiral symmetry that
would be exact for massless quarks is a useful approximate symmetry of
QCD. While the $U(1)\msub{A}$ part is broken by the anomaly, the
$SU(2)\msub{A}$ part is spontaneously broken at low temperatures. The
crossover from the hadronic to the quark-gluon plasma state is signaled by a
large drop of its approximate order parameter, the chiral condensate. The
Banks-Casher relation \cite{Banks:1979yr} connects the chiral condensate to
the spectral density of the quark Dirac operator at zero as
\begin{equation}
   \lim_{m_q\rightarrow 0} \langle \bar{\psi}\psi \rangle \propto \rho(0).
\end{equation}
This suggests that the realization of chiral symmetry is intimately connected
to the lowest part of the Dirac spectrum. Indeed, for a long time it was
believed that in the high temperature phase $\rho(0)$ vanishes in the chiral
limit, signaling the restoration of chiral symmetry. This view, however, was
challenged by lattice QCD studies when a more precise exploration of the
lowest part of the Dirac spectrum became possible, using chirally symmetric
lattice Dirac operators. A study of the overlap Dirac spectrum on quenched
gauge backgrounds found that rather than going to zero, the spectral density
develops a narrow spike at zero \cite{Edwards:1999zm}. The statistics of the
eigenvalues in the spike was shown to be consistent with mixing
instanton--anti-instanton zero modes of a free instanton gas.

This finding was largely ignored for some time, partly because it could be
dismissed as a quenched artifact, the result of using an approximation
omitting the back reaction of the quarks on the gauge field. Indeed, in the
presence of dynamical quarks, each configuration receives in the path integral
an additional weight, proportional to the determinant of the Dirac operator,
and this is expected to suppress the spike of the spectral density at
zero. More recently, detailed studies of the Dirac spectrum with dynamical
quarks on finer lattices revealed that the spike is not a quenched artifact
\cite{Alexandru:2015fxa,Kaczmarek:2021ser,Ding:2020xlj,Alexandru:2023xho}. Some
doubts, however, still remained, as these works used staggered or Wilson sea
quarks that, due to their lack of exact chiral symmetry, cannot precisely
resolve small Dirac eigenvalues, which could lead to an improper suppression
of the spike. Indeed, studies by the JLQCD Collaboration show that for the
proper suppression of the peak, the sea quark action should also be chiral
\cite{Aoki:2020noz}. However, it is possible that even though the peak is
still there, chiral quarks suppress it so much that larger volumes would be
needed to detect it, especially at temperatures farther above the crossover,
where the instanton density is already strongly suppressed by the
temperature. These calculations with chiral fermions are extremely expensive,
and presently cannot be performed on very large volumes. The situation became
even more complicated when the spectral spike was claimed to be singular at
zero, and based on this the presence of a new, intermediate phase of QCD was
suggested \cite{Alexandru:2019gdm}.

Since the Banks-Casher relation contains the spectral density at the singular
point, it is not clear, how the restoration of chiral symmetry occurs. The
fate of the flavor singlet axial symmetry $U(1)\msub{A}$ is especially
interesting, as the question of whether it gets restored at high temperature
has important consequences for the phase structure of QCD-like theories
\cite{Pisarski:1983ms}. In spite of all the efforts in lattice QCD, there is
still an ongoing debate about whether $U(1)\msub{A}$ gets restored
\cite{Kaczmarek:2021ser,Aoki:2020noz}. 

Here, following Ref.~\cite{Edwards:1999zm}, we propose to interpret the
accumulation of small Dirac eigenvalues in the spectral spike, in terms of
instantons. Indeed, in a high statistics quenched lattice QCD study we showed
that the distribution of the topological charge and the number of eigenvalues
in the spike are both consistent with an ideal instanton gas
\cite{Vig:2021oyt}. Meanwhile, the noninteracting nature of the instanton gas
above $T_c$ was independently confirmed
\cite{Bonati:2013tt,Borsanyi:2022fub,Borsanyi:2021gqg}.

In this Letter we provide further evidence that the spike in the spectral
density is due to a free instanton gas, and find that it survives for
arbitrarily high temperatures and nonzero quark masses. We show that in the
thermodynamic limit the spectral density is singular at the origin, and this
singularity dominates the chiral condensate and the $U(1)\msub{A}$ breaking
susceptibility in the chiral limit. This suggests a whole new picture of the
manifestation of chiral symmetry in high temperature QCD, and also solves the
problem of $U(1)\msub{A}$ restoration.

Our results are based on a simple random matrix model of the Dirac operator in
the subspace spanned by instanton zero modes, the so called zero mode zone
(ZMZ). Several versions of this model were used in the past
\cite{Shuryak:1992pi,Schafer:1996wv}. However, our proposal is much simpler
than any previous version, and more importantly, it precisely captures the
details of the spectrum of the lattice overlap Dirac operator on quenched
gauge backgrounds. This not only lends further support to the instanton
explanation of the spectral spike, but also makes it possible to explore its
fate in the chiral limit, a question still impossible to address directly with
state of the art lattice simulation techniques.

{\it The model for quenched QCD.---} We construct a random matrix model,
describing the Dirac operator, restricted to the subspace of instanton zero
modes. In an ideal gas the number of instantons $\nii$ and anti-instantons
$\naa$ are randomly distributed with independent and identical Poisson
distributions of mean $\chi\msub{0} V/2$, where
$\chi\msub{0}= \frac{1}{V} \langle Q^2 \rangle = \frac{1}{V} \langle (\nii -
\naa )^2 \rangle$ is the topological susceptibility and $V$ is the four-volume
of the system.  At high temperatures, the instanton size is limited by the
(Euclidean) temporal size of the system, therefore instantons typically occupy
all the available space in the temporal direction. Thus their location is only
characterized by the three spatial coordinates, which, as the instanton gas is
ideal, are chosen randomly in a 3D box of size $L^3$.

We now construct the random matrix, representing the Dirac operator in the ZMZ
of such an instanton configuration. The size of the matrix is
$( \nii + \naa) \times (\nii +\naa)$, the dimensionality of the ZMZ being
equal to the total number of would be zero modes of the topological
objects. Since the zero modes of same type objects are protected from mixing
by the index theorem, the matrix has two diagonal blocks of zeros of size
$\nii \times \nii$ and $\naa \times \naa$. The rest of the matrix elements
connect instanton and anti-instanton zero modes that are expected to fall off
exponentially with an exponent $\pi T$ \cite{Gross:1980br}. We set the matrix
element, connecting instanton $k$ and anti-instanton $l$ to be
$w_{kl}=iA\cdot e^{-\pi T r_{kl}}$, where $r_{kl}$ is their distance.

The model has two parameters. The topological susceptibility, $\chi\msub{0}$
that sets the instanton density, and $A$. We use a set of $20000$ quenched
$32^3\times 8$ lattice gauge configurations, produced at $T=1.1T_c$ (Wilson
$\beta=6.13$) to fix the parameters. The susceptibility is determined by
counting the number of zero eigenvalues of the lattice overlap Dirac operator,
yielding $\langle Q^2 \rangle=2.58(3)$.  We fit $A$ to reproduce the
distribution of the lowest eigenvalue of the overlap Dirac operator on the
lattice configurations.
    
\begin{figure}[t]
  \centering
  \includegraphics[width=\figsize\textwidth]{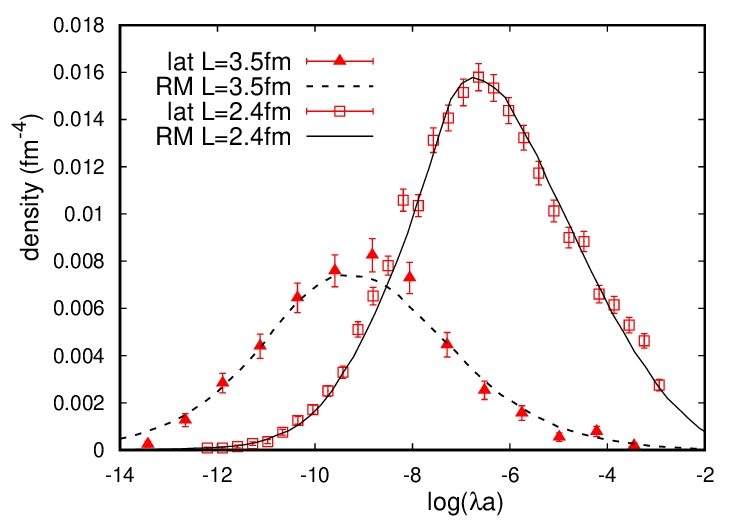}
  \caption{The distribution of the log of the lowest Dirac eigenvalue in two
    different volumes on the lattice and in the random matrix model. The
    smaller volume was used to fit $A$, and for the larger volume the
    prediction of the model is shown, without any further fitting. }
  \label{fig:h0_fitAB}
\end{figure}
   
In Fig.~\ref{fig:h0_fitAB} we show the distribution of the lowest eigenvalue
of the Dirac operator on the lattice and in the random matrix model with the
best fit parameter $A=0.40$ (in lattice units). To better resolve these small
eigenvalues, we plotted the distribution of the natural log of the
eigenvalues. It is already nontrivial that the distribution can be fitted with
just this parameter. A further test is provided by comparing the distribution
of the lowest eigenvalues on different volumes, illustrated in the same
figure. Similar comparisons in different topological sectors, and for the full
spectral density, all show good agreement between the lattice and the random
matrix model. This gives us confidence that the model captures all the
essential features of the lowest part of the lattice overlap Dirac spectrum in
the quenched case. Preliminary data also suggest that the parameter $A$
practically does not depend on the temperature in the range
$1.05T_c$--$1.15 T_c$, whereas $\chi_0$ changes by a factor of 3.

{\it Dynamical quarks.---} On the lattice, including dynamical quarks amounts
to supplementing the quenched Boltzmann weight with the determinant of the
quark Dirac operator, obtained after integrating out the quarks in the path
integral. The determinant can be written as
\begin{equation}
  \det\left( D(A)+m \right)^{N_f} =
  \prod_{ZMZ} (\lambda_i +m)^{N_f} \times \prod_{bulk} (\lambda_i +m)^{N_f},
\end{equation}
where $D(A)$ is the gauge field dependent covariant Dirac operator, and for
simplicity we assume $N_f$ degenerate flavors of quarks with mass $m$. The
product of eigenvalues is split into the lowest eigenvalues (ZMZ) and the rest
(bulk). At high temperatures, the ZMZ and the bulk are well separated in the
spectrum, and are not expected to be correlated. Therefore, if the quark mass
is small, the suppression of the eigenvalues in the ZMZ depends only on the
eigenvalues in the ZMZ. This part of the spectrum is well-described by our
random matrix model, so including the determinant of the random matrices in
the weight will give a self-consistent description of the suppression of the
spectral spike by dynamical quarks.

\begin{figure}[t]
  \centering
  \includegraphics[width=\figsize\textwidth]{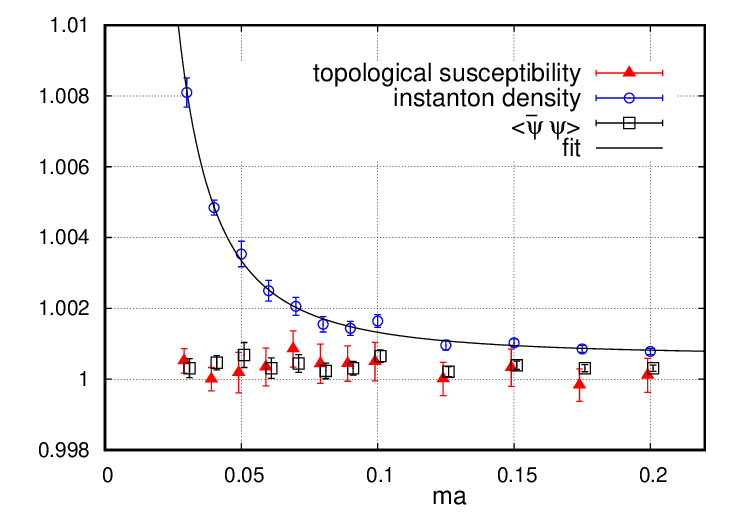}
  \caption{The topological susceptibility and the instanton density normalized
    by $m^2\chi\msub{0}$, and the chiral condensate normalized by
    $m\chi\msub{0}$ as a function of the quark mass. The fit is of the form
    $c/m^2+d$, where $c$ and $d$ are constants.}
  \label{fig:nf2_sim}
\end{figure}

{\it Simulation of the model and results.---} The instanton-based random matrix
model can be easily simulated numerically, even with the additional
determinant factor. For that we used the previously described parameters and
two degenerate quark flavors. (After including dynamical quarks, the
  effective temperature in units of $T_c$ will be higher than the original
  quenched $1.1T_c$, since $T_c$ with dynamical quarks is smaller than in
  the quenched case.) In Fig.~\ref{fig:nf2_sim} we show the topological
susceptibility, the instanton density and the chiral condensate, obtained from
these simulations, as a function of the quark mass. For better visibility, the
first two were divided by $m^2 \chi\msub{0}$, while the chiral condensate was
divided by $m \chi\msub{0}$. We can conclude that to a very good approximation
\begin{equation}
  \chi(m) \approx m^2 \chi\msub{0} \;\;\; \mbox{and} \;\;\;
  \langle \bar{\psi} \psi \rangle \approx m \chi\msub{0}.
  \label{eq:chi_pbp}
\end{equation}
A fit to the instanton density reveals that its behavior in the chiral limit
is well approximated by $m^2\chi\msub{0} + c$.

For the interpretation of the results it is instructive to inspect the
determinant in more detail. On a configuration with $\nii$ instantons and
$\naa$ anti-instantons the random matrix has $|\nii - \naa|$ exact zero
eigenvalues, and the rest of the eigenvalues come in complex conjugate
imaginary pairs. So the determinant can be written as
\begin{equation}
  \det\left( D+m \right)^{N_f} =
  m^{N_f (\nii + \naa )}
  \prod_{\mbox{\tiny pairs} \; i} \left( 1 + \frac{\lambda_i^2}{m^2} \right)^{N_f}. 
\end{equation}
If the eigenvalues $\lambda_i$ are much smaller than the quark mass, each
(anti)instanton contributes with a factor of $m^{N_f}$ to the determinant, and
the distribution of (anti)instanton numbers are still Poissonian, but with a
density suppressed by a factor $m^{N_f}$. This explains the result for the
susceptibility in Eq.~(\ref{eq:chi_pbp}).  One could think that in the
$m\rightarrow 0$ limit, the condition $|\lambda_i| \ll m$ fails to be
satisfied. However, as our simulations reveal, this is not the case, because
as the instanton density drops in the chiral limit, and the typical distance
between instantons grows, the exponentially small matrix elements and, as a
result, the eigenvalues of $D$ become small, and always remain much smaller
than $m$.
   
\begin{figure}[t]
  \centering
  \includegraphics[width=\figsize\textwidth]{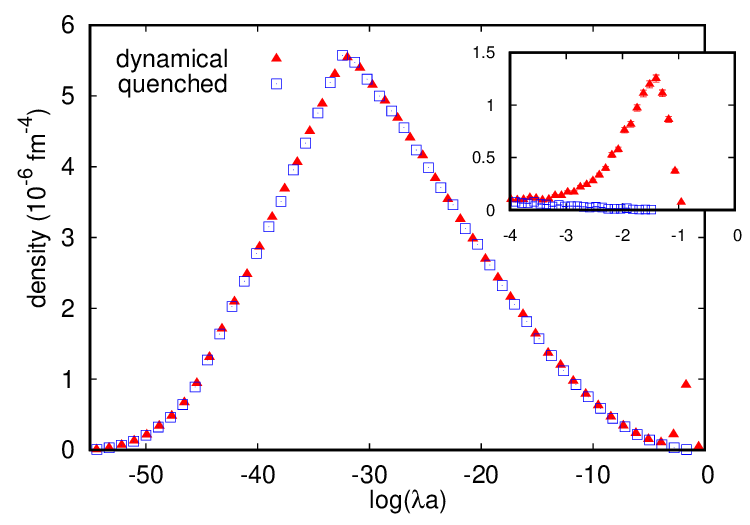}
  \caption{The logarithmic spectral density of the matrix model with two
    flavors of quarks of mass $ma=0.05$ compared to the spectral density of
    the quenched matrix model with the same topological susceptibility. The
    inset is an enlargement of the largest eigenvalues where the two spectra
    significantly differ.}
  \label{fig:spd_m0.05}
\end{figure}

\begin{figure}[t]
  \centering
  \includegraphics[width=\figsize\textwidth]{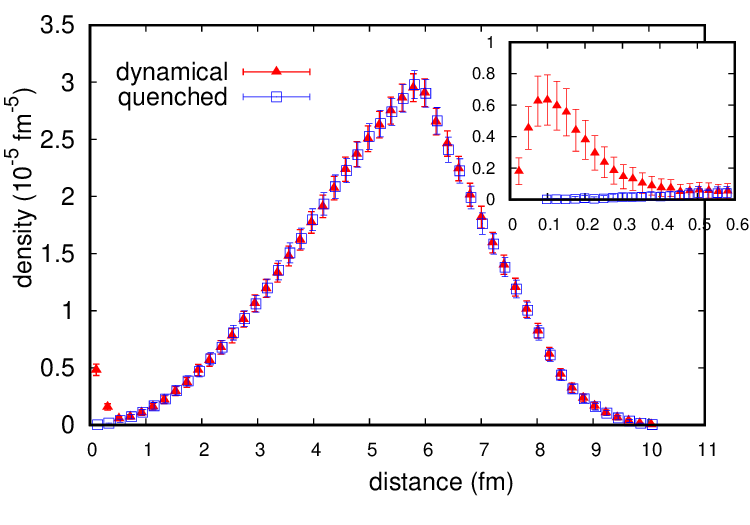}
  \caption{The distribution of the distance of the closest opposite charged
    object on the dynamical and quenched matrix model configurations used for
    Fig.~\ref{fig:spd_m0.05}.} 
  \label{fig:mdia_m0.05}
\end{figure}

To demonstrate that even in the chiral limit, a free instanton gas is
responsible for the spectral spike, in Fig.~\ref{fig:spd_m0.05} we compare the
spectral density of the matrix model with light dynamical quarks with that of
the quenched theory with the same topological susceptibility. The quenched and
the dynamical spectrum fully agree, except for the very largest eigenvalues,
where the dynamical spectrum has an excess of eigenvalues.

These eigenvalues are connected to the component of the instanton density that
is constant in the chiral limit (see Fig.~\ref{fig:nf2_sim}), and appears in
addition to the free instanton gas component, causing the spectral
spike. Further examination reveals that this component of the instanton gas
consists of closely bound instanton--anti-instanton ``molecules''. To show this,
in Fig.~\ref{fig:mdia_m0.05} we plot the distribution of the distance of the
closest opposite charged objects on the dynamical and quenched ensemble. The
two distributions exactly match, except for the shortest distance of
$d \lesssim 1/T$, indicating that the instantons in excess of the free gas,
occur in closely bound molecules.

We emphasize that our random matrix model contains both the gauge action and
the fermion action in a self-consistent way. The gauge action, which, as the
lattice data show, does not produce any interaction among topological
objects, is taken into account by selecting free-instanton configurations with
the parameters given by the quenched lattice data. The fermion action is
included as an additional weight, proportional to the quark determinant for
each configuration. It is only the latter that generates interactions, binding
the molecules.

\begin{figure}[t]
  \centering
  \includegraphics[width=\figsize\textwidth]{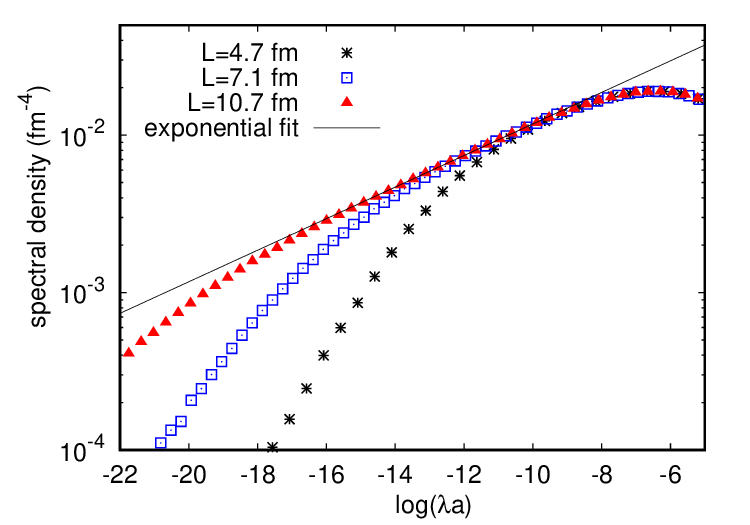}
  \caption{The spectral density (density of $\log \lambda$) for different
    system sizes in the quenched matrix model (parameters given in the
    text). }
  \label{fig:spd_singular}
\end{figure}

{\it Singular spectral density ---} Since a singularity in the spectral
density, suggested in Ref.~\cite{Alexandru:2019gdm} can appear only in the
thermodynamic limit, a precise determination of its power is only possible by
using large spatial volumes. In Fig~\ref{fig:spd_singular} we show the density
of the log of the eigenvalues for three different system sizes in the quenched
matrix model, up to volumes currently not accessible with lattice
simulations. If the spectral density is of the form
$\rho(\lambda) \propto \lambda^\alpha$, then the density of
$\tilde{\lambda}=\log(\lambda)$ goes as
$\tilde{\rho}(\tilde{\lambda}) \propto e^{(1+\alpha)\tilde{\lambda}}$. Fitting
this exponential form to the common envelope of the curves for the different
volumes in the figure, yields $\alpha=-0.769(5)$. A systematic study of the
strength of the singularity as a function of the instanton density is out of
the scope of the present work, but performing simulations with a few different
densities shows that the singularity becomes stronger with decreasing
instanton density. Both the presence of the singularity in the thermodynamic
limit and its strengthening as the instanton gas gets more dilute, can be
qualitatively understood by noting that both large volumes and diluteness
result in smaller matrix elements.

{\it The chiral limit.---} Since as we have already seen, the lowest, singular
part of the spectrum is due to the free-instanton component of the gas which
persists for arbitrarily small quark masses, the singularity will also remain
in the chiral limit. In fact, as $m \to 0$ and the instanton density
decreases, the singularity gets stronger. However, the total weight of the
singularity (its integral) is equal to the density of instantons in the free
component of the gas, and decreases as the topological susceptibility,
$\chi(m) \propto m^{N_f}$. Thus the singularity should remain integrable in
the chiral limit and might eventually become a Dirac delta at zero, as
discussed recently in Ref.~\cite{Azcoiti:2023xvu}.

Let us now discuss the question of chiral symmetry restoration at high
temperature in the chiral limit. With a spectral density singular at the
origin, it is not immediately clear, how to apply the Banks-Casher
formula. However, recalling its derivation, we can write the trace of the
quark propagator in the chiral limit as
\begin{equation}
  \langle \bar{\psi}\psi \rangle \propto
     \langle \sum_i \frac{m}{m^2 + \lambda_i^2} \rangle = 
     \underbrace{\left( \mbox{\parbox{10ex}{\tiny avg.\ number
             of instantons in free gas}}\right) }_{m^{N\msub{f}} \chi\msub{0} V}
     \cdot \frac{1}{m} \; = \; m^{N\msub{f}-1} \chi\msub{0} V.
   \label{eq:pbp}
\end{equation}
Here, we used the fact that eigenvalues, produced by the free instanton
component, satisfy $|\lambda| \ll m$, and each of them contributes a term
$1/m$ to the sum, whereas the eigenvalues corresponding to the molecules have
eventually $|\lambda| \gg m$, and their contribution vanishes as $m/\lambda^2$
in the chiral limit. This explains our simulation result in
Eq.~(\ref{eq:chi_pbp}) for the chiral condensate. It follows that if
$N_f \geq 2$, the order parameter of chiral symmetry breaking vanishes, and,
as expected, the spontaneously broken $SU(N_f)\msub{A}$ is restored, albeit in
a nontrivial way. For $N_f=1$, $\langle \bar{\psi}\psi \rangle$ has a
nonvanishing limit, but in this case it is not an order parameter, as there is
no corresponding symmetry.  In the same fashion, we can also compute the
behavior of the $U(1)\msub{A}$ breaking susceptibility as
\begin{equation}
\chi_\pi - \chi_\delta \propto
     \langle \sum_i \frac{m^2}{(m^2 + \lambda_i^2)^2} \rangle
     \propto  m^{N\msub{f}} \chi\msub{0} V
     \cdot \frac{1}{m^2} \; = \; m^{N\msub{f}-2} \chi\msub{0} V,
   \label{eq:chipd}
\end{equation}
vanishing in the chiral limit only if $N_f \geq 3$, so in the physically
interesting $N_f=2$ case it is nonzero, indicating that the anomalous part of
the chiral symmetry is not restored.

{\it Discussion and conclusions.--- } We emphasize that the temperature enters
our discussion only through the value of the quenched topological
susceptibility, $\chi_0$, and in particular, we only assume that
$\chi_0\neq 0$, which is expected to be true at any finite temperature
\cite{Borsanyi:2015cka}. This implies that our reasoning applies to any large
enough, but finite temperature, and for $N_f=2$ the $U(1)\msub{A}$ part of the
symmetry is not restored at any finite temperature. These results are
consistent with the quasi-instanton picture of Kanazawa and Yamamoto
\cite{Kanazawa:2014cua}, and also with their chiral expansion of the QCD free
energy density \cite{Kanazawa:2015xna}.

The main assumption underlying our instanton-based random matrix model is that
the bulk of the spectrum is not correlated with the ZMZ, and the suppression
of the spectral spike is dominated by the eigenvalues in the ZMZ. This is
certainly true at high enough temperatures, as in this limit the bulk and the
ZMZ get more separated, since the location of the lowest part of the bulk is
controlled by the Matsubara frequency $\pi T$, whereas the spike becomes
narrower in a more dilute instanton gas at higher temperature.

In the explanation of the simulation results we used the property of the ZMZ
that the eigenvalues satisfy $|\lambda| \ll m$, even in the chiral limit. At
higher temperatures the eigenvalues in the ZMZ will become even smaller, and
this condition holds even more precisely. In contrast, for lower temperatures,
toward and beyond $T_c$, the free-instanton picture eventually breaks down,
and our model is not expected to be applicable. Current large-scale overlap
simulations around $T_c$ might be able to tell how exactly this happens
\cite{Fodor:2024pwv}.

We saw that for light quarks, the only deviation of the system from a
free-instanton gas is the appearance of a component of
instanton--anti-instanton molecules of the size $\lesssim 1/T$. The density of
these molecules depends on the details of the instanton--anti-instanton
interaction, which we do not have access to. However, as the quenched spectrum
shows, the gauge interaction is extremely short-ranged, and so is the
fermionic interaction, as our simulations reveal (molecules are smaller than
$1/T$). Therefore, whatever the details of the gauge interaction, it does not
significantly affect the size of the molecules, so the corresponding Dirac
eigenvalues will remain at a fixed scale in the chiral limit, and their
contribution to Banks-Casher type sums of the form Eqs.~(\ref{eq:pbp}) and
(\ref{eq:chipd}) vanishes with some power of the quark mass. It is only the
singular part of the spectrum, produced by the free-instanton gas that
controls whether these quantities are nonvanishing in the chiral limit.

%%%%%

\end{document}